\documentstyle[12pt]{ioplppt}

\begin{document}

\sloppy

\jl{2}

\paper{Influence of perturbations on the electron wave function
inside the nucleus}

\author{M. Yu. Kuchiev\ftnote{1}{E-mail: kuchiev@newt.phys.unsw.edu.au}
and V. V. Flambaum\ftnote{2} {E-mail: flambaum@newt.phys.unsw.edu.au} }

\address{School of Physics, University of New South Wales,
Sydney 2052, Australia}

%**************************************************************************
\begin{abstract}
A variation of the valence electron wave function inside a nucleus induced
by a perturbative potential is expressed in terms of the potential momenta.
As an application we consider QED  vacuum polarization 
corrections due to the Uehling and Wichmann-Kroll potentials
to the  weak interaction matrix elements.

\end{abstract}

%Submitted to J. Phys. B: At. Mol. Opt. Phys. (2001)


\maketitle

\section{Introduction}\label{intro}

Precise low-energy experiments on parity nonconservation (PNC) in atoms 
provide a test of the standard model of elementary particle interactions. 
In a recent PNC experiment with cesium by Wood {\it et al}
\cite{wood97} the 
PNC E1 transition amplitude between the $6s$ and $7s$ states has been 
determined with an unprecedented accuracy of 0.3 \%.
At this level of accuracy a small perturbative potential
(such as the Breit interaction, QED vacuum polarization) may influence
the result. The PNC weak interaction matrix elements are determined
by the electron wave function inside a nucleus.
Having this, as well as some other possible applications, in mind 
we present  in this paper a simple analytical approach which allows one
to calculate corrections to the electron wave function
near the origin due to a perturbative potential. As an application
we consider corrections to weak matrix elements produced
by the QED vacuum polarization (Uehling potential \cite{uehling},
Wichmann-Kroll potential \cite{wichmann-kroll}) and a QED correction
to the electron potential.
Relativistic units $c=\hbar = 1,~e^2 = \alpha=1/137.04,~\alpha a_0 =
m^{-1}$ ($a_0$ is the Bohr radius) are used below everywhere, 
if not specified otherwise..

\section{General formalism}
\label{general}
Our goal is to describe
a variation of an electron wave function inside the nucleus
due to some local potential $V(r)$ considered as a perturbation.
Let us presume that the electron motion in an atom is described
with the help of a single-electron spherically symmetrical Hamiltonian $H$,
which is, generally speaking, relativistic, i.e. of the Dirac type.
The wave function for
the $n$-th energy level $\psi_{njl}({\bf r})$
which is the Dirac $4$-spinor
characterized by the total momentum $j$, orbital momentum $l$ and
projection of the total momentum $\mu$, the latter index is suppressed,
satisfies the Schroedinger equation
\begin{equation}\label{schr}
E_{njl} \psi_{njl}({\bf r}) = H \psi_{njl}({\bf r})~,
\end{equation}
with appropriate boundary conditions at $r=0$ and $\infty$.  The
spherically symmetrical potential $V(r)$ considered
as a perturbation results in a variation of the wave function $\delta
\psi_{njl}(r)$ that we are looking for.
This variation can obviously be presented as
\begin{equation}\label{delta}
\delta \psi_{njl}({\bf r}) = \int \tilde G_{jl}({\bf r},{\bf
r}';E_n)\,V(r)\, \psi_{njl}({\bf r}')\,{\rm d} ^3 r'~,
\end{equation}
where $\tilde G_{jl}({\bf r},{\bf r}'; E_n)$ is the corresponding Green
function for the operator $H$, that
can be expanded as a series over a full
set of solutions of the Schroedinger equation (\ref{schr}) for the $j,l$
wave
\begin{equation}\label{green}
\tilde G_{jl}( {\bf r},{\bf r}'; E_n ) =
\sum_{n' \ne n}  \frac{\psi_{n'jl}({\bf r})\psi_{n'jl}^+({\bf
r}')}{ E_n-E_{n'} }~.
\end{equation}
This Green function is necessarily a
$4\times 4$ matrix in the Dirac indices that are suppressed.
Remember that
the $n$-th electron level is excluded from summation in (\ref{green}). To
keep trace of this fact a tilde hat is used for $\tilde G_{jl}({\bf r},{\bf
r}';E_n)$ thus distinguishing it from the Green function $G_{jl}({\bf
r},{\bf r}',E)$ defined for arbitrary, non-specified energy $E$.  Using
orthogonality and a completeness of a full set of the wave functions
$\psi_{njl}({\bf r})$ that satisfy  the Schroedinger equation (\ref{schr}),
i.e. the fact that
\begin{eqnarray}\label{norm}
&&\int \psi_{njl}^+({\bf r})\psi_{n'jl}({\bf r})\,{\rm d} ^3r =\delta_{nn'}~, \\
\label{compl}
&&\sum_{n,j,l,\mu} \, \psi_{njl}^+({\bf r}) \psi_{njl}({\bf
r}') = \delta({\bf r}-{\bf r}')~,
\end{eqnarray}
one finds from
(\ref{green}) that $\tilde G_{jl}({\bf r},{\bf r}';E_n)$ satisfies the
following non-homogeneous equation
\begin{equation}\label{greensch} \left(
E_n -H \right) \tilde G_{jl}({\bf r},{\bf r}';E_n) = \delta({\bf r}-{\bf
r}')- \psi_{njl}({\bf r})\psi_{njl}^+({\bf r}')~, \end{equation}
as well as an integral condition
\begin{equation}\label{ortho} \int \tilde
G_{jl}({\bf r},{\bf r}';E_n) \psi_{njl}({\bf r}') {\rm d} ^3r = 0~.  \end{equation}
The last term in the right-hand side of (\ref{greensch}) as well as
condition
(\ref{ortho}) arise due to the same, mentioned above reason, namely
that the $n$-th energy level is excluded from summation in (\ref{green}).

Our goal is to describe the wave function variation inside the nucleus.
To this end it is sufficient to find its variation
only at one point inside, for example  at the nuclear
center which we consider as an origin.
Having done that one recovers the wave function everywhere
inside (and in close vicinity of) the nucleus simply by scaling its
non-perturbed value to comply with the variation found at the chosen point.
This statement follows from the fact that the potentials
considered in this paper are supposed to produce negligible effects
inside the nucleus. They contribute only due to their
existence in a nuclear exterior region.  This means that equations
governing the electron wave function inside the nucleus remain intact by
the perturbation.  As a result the perturbation can only scale the wave
function inside. We will use this fact looking for the wave
function specifically at the origin, i.e. hunting for $\delta
\psi_{njl}({\bf 0})$.

In order to find the wave function at the origin
one needs to find the Green function $\tilde G({\bf 0},{\bf r}';E_n)$.
This task can be conveniently
fulfilled using the following method. Consider
an energy $E$ as an arbitrary parameter assuming only that
$E$ is located
in some vicinity of a chosen atomic energy level
$E_n$. Let us call  $\psi_{jl}({\bf r},E)$ a solution
of the Schroedinger equation
\begin{equation}\label{schrE}
(E-H)\psi_{jl}({\bf r},E) = 0~.
\end{equation}
Obviously for an arbitrary energy $E$ this solution cannot satisfy proper
boundary conditions both at infinity and at
the origin. However, we can always consider
the proper condition at one of the
two points. We may assume therefore that $\psi_{jl}({\bf r},E)$
decreases at infinity
\begin{equation}\label{infinity}
\psi_{jl}({\bf r},E) \rightarrow 0~, ~~~~r\rightarrow \infty~.
\end{equation}
Solutions of this type are necessarily singular at the origin, if the
energy parameter $E$ does not coincide with some physical energy level.
It is convenient, nevertheless, to introduce the following normalization
integral

\begin{equation}\label{norma} \int 
\psi_{njl}^+({\bf r}) \psi_{jl}({\bf r},E)\, {\rm d} ^3r = 1~.  
\end{equation}
which converges well because the mentioned singularity at the origin is 
compensated for by a zero-type behaviour of the regular solution.

The function $\psi({\bf r},E)$ possesses several important for us
properties.  Firstly, at the point $E=E_n$ the boundary condition at the
origin can obviously be satisfied. Combining this statement with
equations (\ref{schrE}),(\ref{infinity}) and
(\ref{norma})  we conclude that for $E=E_n$ the
function $\psi_{jl}({\bf r},E)$ coincides with the wave function for the
$n$-th energy level
\begin{equation}\label{psipsi} \psi_{jl}({\bf r},E_n) =
\psi_{njl}({\bf r})~.  \end{equation}
To reveal another useful property of
$\psi_{jl}({\bf r},E)$, let us differentiate the Schroedinger equation
(\ref{schr}) over the energy at the point $E=E_n$
\begin{equation}\label{diffE}
(E_n-H)\frac{\partial \psi_{jl}({\bf r},E_n)}{\partial E}
= - \psi_{jl}({\bf r},E_n)~, ~~~r > 0~.
\end{equation}
Keeping in mind that $\psi_{jl}({\bf r},E)$ is irregular at
the origin we have to suspect that the right-hand side of (\ref{diffE}) may
include the delta-term $\propto \delta ({\bf r})$, or its derivatives. This
suspicion, justified below, prompts to remember that $r>0$ when
(\ref{diffE}) is taken literally.

To proceed let us differentiate the normalization condition
(\ref{norma}) over $E$ at the point $E=E_n$
\begin{equation}\label{diffnorma}
\int \psi_{njl}^+({\bf r})
\frac{\partial \psi_{jl}({\bf r},E_n)}{\partial E}\,{\rm d} ^3r = 0~.
\end{equation}
Compare now a set of equations (\ref{diffE})
and (\ref{diffnorma}) with (\ref{greensch}) taken at $r=0, ~r'> 0$ and
(\ref{ortho}).  Their obvious identity indicates that the Green function at
the origin $\tilde G_{jl}({\bf 0},{\bf r};E_n)$ can be presented as
\begin{equation}\label{Gpsi}
\tilde G_{jl}({\bf 0},{\bf r};E_n)
= \psi_{njl}({\bf 0}) \frac{\partial \psi_{jl}^+({\bf r},E_n)}
{\partial E}~.
\end{equation}
We can now clarify behaviour of the right-hand
side of equation (\ref{diffE}) at $r=0$.  From (\ref{Gpsi}) and
(\ref{greensch}) we deduce that our expectation was correct, at $r=0$ the
right-hand side of (\ref{diffE}) should indeed be modified to include an
additional delta-term $\sim \delta({\bf r})$ (though for our purposes
it suffices to consider this equation only at $r>0$).

We find that $\psi_{jl}({\bf r},E)$ is a very convenient object.
It allows one to describe simultaneously a set of wave
functions $\psi_{njl}({\bf r})$ (\ref{psipsi})
as well as the Green function $\tilde G({\bf 0},{\bf r};E_n)$ (\ref{Gpsi}).
It follows from (\ref{delta}),(\ref{Gpsi}) that the variation
of the wave function due to perturbation $V(r)$ can be expressed as
\begin{equation}\label{psi0}
\delta \psi_{njl}({\bf 0}) = \psi_{njl}({\bf 0})\,
\int \frac{\partial \psi_{jl}^+({\bf r},E_n) }{\partial E} V(r)
\psi_{njl}({\bf r})\,{\rm d} ^3r~.
\end{equation}
This convenient presentation is the main result of this Section.
Its applications are discussed below.

\section{Perturbation at small distances}
\label{releff}
Let us consider perturbative potentials $V(r)$
located at small separations from the nucleus.
We will assume, however, that a potential considered
gives a significant contribution mainly outside the
atomic nucleus, i.e. assume that a region of distances where the potential
is important satisfies
\begin{equation}\label{r<1/Z} r_N \ll r \ll a_0/Z~,
\end{equation}
where $r_N$ is the nuclear radius.
 A wave function variation in the nuclear
interior can be described by a scaling factor, which we find below.

Conventional presentation for the Dirac four-spinor
for spherically symmetrical potentials reads
 \begin{equation}\label{spinor}
 \psi_{njl}({\bf r}) =\frac{1}{r}
 \left(
  \begin{array}{cc} f_n(r)\Omega_{jl\mu}({\bf n}) \\
{\rm i} g_n(r)\Omega_{j\tilde l \mu}({\bf n})
  \end{array} \right)~.
\end{equation}
  Here $f_n(r)$ and $g_n(r)$ are
  the large and small radial  components of the
  spinor, the indices $jl$ for them are suppressed. They are
normalized as
  \begin{equation}\label{fgnorma}
  \int_0^\infty (\,f_n^2(r)+g_n^2(r)\,)\,{\rm d}r =1~,
  \end{equation}
  $\Omega_{j l \mu}({\bf n})$ and $\Omega_{j\tilde l \mu}({\bf n})=
-({\mbox {\boldmath $\sigma$}}\cdot {\bf n})$ $\Omega_{j l m}({\bf n})$
  are spherical spinors, and $l+\tilde l =2 j$.
  The Dirac equation (\ref{schr}) for a motion in the potential $U(r)$
  in this notation takes the familiar form
  \begin{eqnarray}\label{dirac}
  && f_n'(r) + \frac{\kappa}{r}f_m(r)
  = (m+E_n- U(r)) \,g_n(r) \\ \nonumber
  - && g_n'(r) + \frac{\kappa}{r}g_n(r) = (m-E_n-U(r)) \, f_n(r) ~,
  \end{eqnarray}
where  $\kappa = l(l+1)-j(j+1)-1/4 = \pm (j+1/2)$.
 For small separations of an electron from the atomic nucleus (\ref{r<1/Z})
  the potential $U(r)$
 can be approximated by the pure Coulomb potential created by the nuclear
 charge $Z$
 \begin{equation}\label{UC} U(r) \simeq - \frac{Z e^2}{r}~.
 \end{equation}
 Additional simplification for this region
 comes from the fact that the energy
 of a valence electron
 is low and therefore for small separations (\ref{r<1/Z})
 one can safely assume that $E \simeq m$.
 With these
 simplifications the Dirac equation (\ref{dirac}) reads
 \begin{eqnarray}\label{diracC}
&& f_n'(r) + \frac{\kappa}{r}f_n(r) =
 \left( 2m + \frac{Ze^2}{r} \right) g_n(r) \\ \nonumber
 - && g_n'(r) +  \frac{\kappa}{r}g_n(r) =
  \frac{Ze^2}{r} f_n(r) ~.  \end{eqnarray}
 There are two sets of solutions for these equations.
 One of them is regular at the origin. From
 (\ref{diracC}) one finds that it behaves as
 \begin{eqnarray}\label{fg}
 f_+(r) &=& a_+ r^\gamma~, \\ \nonumber
 g_+(g) &=& b_+\,r^\gamma ,~~~~~~~~~~~~~
 b_+=\frac{\gamma+\kappa}{Z\alpha}~.  \end{eqnarray}
 Here
 $\gamma = \big(\kappa^2 - (Z\alpha)^2\big)^{1/2}$.  
Clearly, this regular solution
 can be used to describe the electron wave function for small distances
 (\ref{r<1/Z}):  $f_n(r) = f_+(r),~~~g_n(r) = g_+(r)$.  The subscript $+$
 is used to distinguish this solution from the singular one. The
 latter, that will be called $f_-(r),g_-(r)$, behaves as
 \begin{eqnarray}\label{fg-} f_-(r) &=& a_-r^{-\gamma}~, \\ \nonumber
 g_-(g) &=& b_-r^{-\gamma} ,~~~~~~~~~~~~~
b_-=\frac{-\gamma+\kappa}{Z\alpha}~.  \end{eqnarray} This
 solution  is used below for the description of the Green function.  The
 explicit form for both sets of solutions found from (\ref{diracC}) reads,
\begin{eqnarray}\label{besself} f_{\pm}(r) &=& a_{\pm}
 \,\frac{\Gamma(\pm 2\gamma +1)}{(8Z\alpha m)^{\pm\gamma} }
\left[\,J_{\pm 2\gamma}(x)  
- \frac{1}{2(\pm \gamma - \kappa)}\,x J_{\pm 2 \gamma+1}(x) \,\right]
\\ \label{besselg}
%&=& a_{\pm} \,[- \,x J_{\pm 2 \gamma - 1}(x) + 2(\pm \gamma
%              + \kappa )\,J_{\pm 2\gamma}(x)\,]~, \\ \label{besselg}
g_{\pm}(r) &=&  a_{\pm}\,
\frac{\pm \gamma +\kappa}{Z\alpha}\, 
\frac{\Gamma(\pm 2\gamma +1)}{  (8Z\alpha m)^{\pm\gamma} }\,
J_{\pm 2\gamma}(x)~.
\end{eqnarray}
Here $J_\nu(x)$ is the Bessel function,
$x = \sqrt{8 Z\alpha mr}$. 
Numerical coeffici
ents in (\ref{besself}),(\ref{besselg}) are chosen
to satisfy (\ref{fg}),(\ref{fg-}).
The coefficient $a_+$ for the
regular solution that represents the electron wave function should
be found from the normalization condition for this function.
A proper normalization of the
coefficient $a_-$ is discussed in detail below, see (\ref{aa}).

 Following the approach of Section \ref{general} we need to replace
 $E_n$ in (\ref{dirac}) by an arbitrary value $E$, assuming that
 the corresponding solution $\psi_{jl}({\bf r},E)$
 with components $f(r,E),~g(r,E)$ behaves
 regularly at infinity as specified in (\ref{infinity}).
 Further, we need to consider derivatives over the energy
 $\partial f(r,E)/\partial E \equiv f_E(r,E)$ and
 $\partial g(r,E)/\partial E \equiv g_E(r,E)$.
 Our task is to find these functions
 at small $r$.
 Since $f(r,E),~g(r,E)$ are singular at the origin, we have
 to expect that $f_E(r,E_n),~g_E(r,E_n)$ are singular as well.
 Bearing this in mind we can neglect the regular, and therefore small,
 right-hand side of
 (\ref{diffE}) when $r\rightarrow 0$.
 We deduce from this that for small $r$ the functions
 $f_E(r,E_n),~g_E(r,E_n)$ satisfy the homogeneous Dirac equation
 (\ref{dirac}) behaving
singularly in the vicinity of $r=0$. The
notation $f_-(r),~g_-(r)$ introduced above
specifies exactly this solution
of the Dirac equation. It follows from (\ref{fg-}) that for small
separations the following asymptotic conditions hold
\begin{eqnarray}\label{fEg}
f_E(r,E_n) &=& f_-(r) = a_-\,r^{-\gamma}~, \\ \nonumber
g_E(r,E_n) &=& g_-(r) = b_-\,r^{-\gamma}~.
\end{eqnarray}
We need to continue this line of argumentation and find corrections
of the order of $\sim mr$ to the right-hand sides of (\ref{fEg}).  Observe
firstly that according to (\ref{fg}) the right-hand side of non-homogeneous
equation (\ref{diffE}) is small enough for short distances to produce no
corrections of the order of $mr$.  Therefore the main correction to
(\ref{fEg}) arises from the homogeneous Dirac equation (\ref{diracC}) when
the mass term in the right-hand side of this equation is taken into
account.  Technically the easiest way to recover the correction due to the
mass term is through an expansion of the explicit solutions
(\ref{besself}),(\ref{besselg}) in powers of $mr$.
We will present the result below, in (\ref{final}), where
a similar correction for the regular solution (\ref{fg}) is also included.

To proceed  we need to rewrite (\ref{psi0}) in terms
of large and small components of the Dirac spinor. Relations
(\ref{fg}) ensure that variations of both components
due to perturbation
are proportional at the origin. This means that if we define
the component ratios at the origin as a limit
$ \delta f_+(0)/f_+(0) \equiv [\delta f_+(r)/f_+(r)]_{r\rightarrow 0}$
and
$ \delta g_+(0)/g_+(0) \equiv [\delta g_+(r)/g_+(r)]_{r\rightarrow 0}$
(the limit
is necessary since one of the components may turn zero at the origin)
then
\begin{equation}\label{fg0}
 \delta f_+(0)/f_+(0) = \delta g_+(0)/g_+(0)~.
\end{equation}
Using this fact we derive from (\ref{psi0})
\begin{eqnarray}\label{ffgg}
\frac{\delta f_+(0)}{f_+(0)}&=&\frac{\delta g_+(0)}{g_+(0)}
=\int_0^\infty [\,f_n(r)f_E(r,E_n)+g_n(r)g_E(r,E_n) \,]V(r)\,{\rm d}r
\\ \label{ffgg2}
&=&\int_0^\infty [\,f_+(r)f_-(r)+g_+(r)g_-(r)\,]\,V(r)\,{\rm d}r~.
\end{eqnarray}
The last identity here arises because, as  explained above,
for small separations satisfying (\ref{r<1/Z}) the wave function
$f_n(r),~g_n(r)$
and derivatives over energy $f_E(r,E_n),~g_E(r,E_n)$
can be replaced by the {\em plus} and {\em minus} solutions
$f_\pm(r),~g_\pm(r)$ respectively. The magnitude of these latter solutions
is governed by coefficients $a_\pm$ in asymptotic
formulae (\ref{fEg}).
We need therefore to find the product $a_+a_-$ .
This can be achieved using the following transformation.
Multiply the non-homogeneous equation
(\ref{diffE})
by $\psi_{njl}^+({\bf r})$ and integrate over
the full 3D space from which the interior of a sphere
$S_{\varepsilon}$ of radius ${\varepsilon}$
around the origin is excluded. Consider
${\varepsilon}>0$ as a small parameter which is to be put to zero
at the end of the calculations, $\varepsilon \rightarrow 0$.
The normalization condition for $\psi_n({\bf r})$ ensures that
this procedure gives $-1$
in the right-hand side of (\ref{diffE}).
Integrating the Dirac Hamiltonian $H$ in the left-hand side of this equation
by parts one observes
that only the surface term
sitting on the sphere $S_{\varepsilon}$ survives
\begin{eqnarray}\nonumber
\!\!\!\!\!\!\!\!\!\!\!\!\!\!\!\!
\int_{\varepsilon \le r}\psi_{njl}^+({\bf r})(E-H)
\frac{ \partial \psi_{jl}({\bf r},E)}{\partial E}\,{\rm d} ^3r=
 \int_{S_\varepsilon}
\psi_{njl}^+({\bf r}) \,(-{\rm i} {\mbox {\boldmath $\alpha$} }\cdot{\bf n} )\,
\frac{ \partial \psi_{njl}({\bf r},E_n)}{\partial E} \,{\rm d}S
\\ \nonumber
\!\!\!\!\!\!\!\!\!\!\!\!\!\!\!\!
=
- \left(\,f_n(\varepsilon) \frac{ \partial g(\varepsilon,E_n)}{\partial E}-
        g_n(\varepsilon) \frac{ \partial f(\varepsilon,E_n)}
{\partial E}\,\right)
%\\ \nonumber
%\!\!\!\!\!\!\!\!\!\!\!\!\!\!\!\!
=-\left(\,f_+(\varepsilon) g_-(\varepsilon) -
g_+(\varepsilon)f_-(\varepsilon)\,\right)
\\ \label{surf}
\!\!\!\!\!\!\!\!\!\!\!\!\!\!\!\!
=-(\,a_+b_- - b_+a_-)~.
\end{eqnarray}
Here we use representation of the spinors
$\psi_{njl}({\bf r})$ and $\partial \psi({\bf r},E)/\partial E$
in terms of their large and small
components, compare (\ref{spinor}), as well as
the fact that infinity
$r=\infty$ gives no contribution to the surface term since
$\psi_{njl}({\bf r})$ and $\partial \psi_{jl}({\bf r},E)/\partial E$
are regular there.  For small radius
$r=\varepsilon$ one can express $\psi_{njl}({\bf r})$
and $\partial \psi_{jl}({\bf r},E)/\partial E$ in
terms of $f_\pm(r),g_\pm(r)$, as was explained above,
and use asymptotic formulae
(\ref{fg}),(\ref{fg-}), expressing thus the surface term via the
coefficients $a_\pm,b_\pm$ in the last identity in (\ref{surf}).
We find from these transformations that $a_+b_- - b_+a_- = 1$, or,
remembering expressions of $b_\pm$ in terms
of $a_\pm$ in (\ref{fg}),(\ref{fg-}), find the product in question
\begin{equation}\label{aa}
a_+a_- = \frac{1}{2}\,\frac{Z\alpha}{\gamma}~.  \end{equation}
We possess
now all ingredients necessary to derive the final result.  Take equation
(\ref{ffgg2}). Substitute in its right-hand side expressions
(\ref{besself}) and (\ref{besselg}) for $f_\pm(r),g_\pm(r)$ that are
supplemented by condition (\ref{aa}) on the coefficients $a_\pm$.
After that expand the resulting integrand that arises from
the right-hand side of (\ref{ffgg2}) in powers of
$mr$. This expansion is both justified and necessary in view of
the following reasons.  The expansion is allowed because the perturbation
$V(r)$ is located in the region of small separations (\ref{r<1/Z}).
A typical radius where the potential is located is a parameter
for this expansion.
Necessity for this expansion is twofold. Firstly, the approach
developed neglects the screening of the Coulomb field by an electron cloud,
which is a good approximation only in the close vicinity of the nucleus.
Secondly, the procedure described neglects the regular solution in
the right-hand side of the non-homogeneous Dirac equation, which is
justified only for small distances where this solution is small.

Analytical calculations described above are straightforward.
The final result reads 

\begin{equation}\label{final}
\frac{\delta f_+(0)}{f_+(0)}=\frac{\delta g_+(0)}{g_+(0)}=
-\frac{m}{\hbar^2}\,\int_0^\infty V(r)\,(a+kr)\,{\rm d}r~.
\end{equation}
Here $a$ is a parameter with length dimension
while $k$ is a dimensionless coefficient
\begin{eqnarray} \label{a}
a &=&  \frac{Z\alpha}{\gamma}\,\frac{\hbar}{mc}~,
\\ \label{k}
k &=&  \frac{2\kappa(2\kappa-1)}{\gamma(4\gamma^2-1)}~.
\end{eqnarray}
Relations (\ref{final}),(\ref{a}) are presented in absolute units,
to make them more accessible for different applications.

Simple formula (\ref{final}) is one of the most important results
of this paper. It solves
the main problem formulated in this section presenting a variation of the
wave function at the origin in very transparent terms,
as a linear
combination of the zeroth and first momenta of the perturbative potential.

Note that
for the short-range potentials relative corrections to the energy
and wave function are quite different. Indeed, we can
approximate the energy variation in this case using asymptotic
relations (\ref{fg}) as
 \begin{equation}\label{deltaEshort}
 \delta E_n = \langle njl |V|njl \rangle \simeq
 (a_+^2+b_+^2)\int_0^\infty V(r) r^{2\gamma} \,{\rm d}r~.
 \end{equation}
 For an arbitrary perturbation $V(r)$
 the integral in the right-hand side of this identity
 may deviate significantly from the integral in (\ref{final}).
 Therefore for short-range perturbations the
 energy variation, generally speaking,  cannot serve as estimate what
 happens with the wave function.

 An interesting comparison can be made
 with the nonrelativistic limit of (\ref{final}) that reads
 \begin{equation}\label{nonrelat}
 \frac{ \delta \psi_{nl}( 0 ) }{ \psi_{nl}(0) } =
-\frac{m}{\hbar^2} \,\frac{2}{2l+1}\int_0^\infty V(r)\,r\,{\rm d}r~.
 \end{equation}
 Deriving this identity
 we use the fact that according to (\ref{a}) the parameter $a$ turns zero in
 the limit $Z\alpha \rightarrow 0$, while from (\ref{k}) one
 derives $k \rightarrow2/(2l+1)$.
 There is a simple short cut derivation
 that leads to (\ref{nonrelat}) and
 can be used for verification of this result.  Assume that in the
 nonrelativistic limit the electron motion 
 is dominated by the kinetic term, which is true for short
 separations.  Derive from this an approximation
 \begin{equation}\label{G0}
 \tilde G_l({\bf r},{\bf r}';E_n) \simeq G^{(0)}_l
({\bf r}-{\bf r}';E_n) \simeq-\frac{2m}{\hbar^2}
 \frac{1}{2l+1} \frac{r_<^l}{r_>^{l+1}}~,
\end{equation}
where $G^{(0)}_l({\bf r},{\bf r}';E_n)$
 is the Green function for
 the free motion in the $l$-th partial wave and
 the last identity takes into account the fact that
 the binding energy is negligible for short
 distances.
 Remembering also that for nonrelativistic motion the wave function
 behaves as
 $\psi_{l}(r) \propto r^l,~r \rightarrow 0$
 one immediately derives (\ref{nonrelat}) directly from (\ref{delta}) thus
 verifying relativistic equation (\ref{final}) that
 we used above. To comply with absolute units used in (\ref{final})
we use the same units in nonrelativistic formulae
(\ref{nonrelat}),(\ref{G0}).

The nonrelativistic result (\ref{nonrelat}) shows that
the parameter that governs variation of the wave function is
$m\int V(r)r{\rm d}r$. This is almost an obvious result valid for a variety
of quantum mechanical problems \cite{LLIII}.
The relativistic result (\ref{final}) shows that there
exists another parameter $ma\int V(r){\rm d}r$.  
It is suppressed compared with
the nonrelativistic parameter only by a factor $Z\alpha$ which is not small
for heavy atoms.  This suppression can be well compensated for if
the potential considered increases at small separations which makes
$a \int V(r){\rm d}r$
larger than $\int V(r)r{\rm d}r$. In this case the found relativistic
parameter becomes more important than the nonrelativistic one.  A
perturbation due to the QED vacuum polarization discussed in Section
\ref{vacuum} presents an example important for applications.

We can apply the results obtained above for a specific interesting case.
Consider the
parity-violating weak interaction of an atomic electron with the nucleus
that mixes $s_{1/2}$ and $p_{1/2}$ states of an outer electron.  The matrix
element for this mixing $\langle p_{1/2}|W|s_{1/2}\rangle $ is saturated
inside the nucleus. Therefore the  variation 
of the matrix element for the weak interaction can be found 
simply by adding variations of $s_{1/2}$ and $p_{1/2}$ states given in
(\ref{finalu2})

 \begin{equation}\label{weak}
\!\!\!\!\!\!\!\!\!\!
\frac{\delta
 \langle p_{1/2}|W|s_{1/2}\rangle} {\langle p_{1/2}|W|s_{1/2}\rangle} =
-\frac{2m}{\hbar^2}\,\int_0^\infty V(r)\,\left(a+\frac{2}{3} kr\right)\,{\rm d}r~.
 \end{equation}
Deriving this result 
we take into account that essential parameters for $s_{1/2}$
 and $p_{1/2}$ states are $\kappa_{s}=-1,~\kappa_{p}=1, ~\gamma_s = \gamma_p
 = (1-(Z\alpha)^2)^{1/2}\equiv \gamma$ and assumed that $k\equiv k_s=
6/\Big( \gamma(4\gamma^2-1) \Big)$.

\section{Vacuum polarization}
\label{vacuum}
 Let us apply  (\ref{final})
 to a specific case when perturbation originates from
 polarization of the QED vacuum caused by the Coulomb field of the
 nucleus.  In the lowest, second order of the QED perturbation theory this
 polarization is described by the  Uehling potential  \cite{uehling}
 $V_{\rm VP}(r)$
 \begin{eqnarray}\label{uehling}
V_{\rm VP}(r) &=&
-\frac{2\alpha}{3\pi} \,\left(\frac{Ze^2}{r}\right)\, \int_1^\infty
 \exp(-2mr\zeta)\,Y(\zeta)\,{\rm d}\zeta~, \\ \label{Y}
 Y(\zeta) &=& \left(1+\frac{1}{2\zeta^2}\right)
\frac{ \sqrt{ \zeta^2-1}}{\zeta^2}~.
 \end{eqnarray}
The Uehling potential (\ref{uehling}) is singular at the origin
\begin{equation}\label{uo}
V_{\rm VP}(r) = - \frac{2 \alpha}{3\pi} \,
 \left( \frac{Ze^2}{r} \right) \,
 \left( \ln\frac{1}{mr}-C-\frac{5}{6}\right), ~~~~~mr\ll 1~.
 \end{equation}
Here $C = 0.577\dots$ is the Euler constant.
A $\ln mr $ function in (\ref{uo}) describes conventional
 scaling of the QED coupling constant $e^2$ that
 manifests itself for short distances. This scaling factor
has an interesting consequence for the problem at hand.
Being introduced in (\ref{final}) it results in the $\ln^2 mr$
divergence of the integral
$ma\int V_{\rm VP}(r){\rm d}r$  at
small $r$. This divergence is eliminated by the finite nuclear
size. As a result we find an estimate for the variation
of  the weak matrix element (\ref{weak}) 
\begin{equation}\label{log2}
\frac{\langle p_{1/2}|W|s_{1/2}\rangle} {\langle p_{1/2}|W|s_{1/2}\rangle} 
\sim
- 2 ma\int V_{\rm VP}(r){\rm d}r \sim  
\frac{2}{3\pi\gamma}\,\alpha(Z\alpha)^2\,\ln^2(mr_N)~.
\end{equation}
This shows that there exists the $\ln ^2 mr_N$ enhancement in the
problem, as was firstly discovered by Milstein and Sushkov 
\cite{milstein-sushkov-01} using
other methods. In our approach this result 
is linked with the relativistic
parameter $ma\int V(r){\rm d}r$ (\ref{a}) 
introduced in Section \ref{releff}.
For heavy atoms the $\ln^2 mr_N$ enhancement
compensates for the additional
suppressing factor $Z\alpha$ in the
relativistic parameter $a$ (\ref{a}) thus making this parameter
dominant.

In order to present more accurate results describing the influence
of the vacuum polarization on the wave function let us
substitute (\ref{uehling}) in (\ref{final}). Integrating over $r$
we find
\begin{eqnarray} \nonumber
%\label{finalu}
\frac{\delta f_+(0)}{f_+(0)}&=&\frac{\delta g_+(0)}{g_+(0)} =
  \frac{2 Z \alpha^2}{3\pi}
 \int_1^\infty \left(\, am E_1(mr_N \zeta) + \frac{k}{2\zeta} \right)
 Y(\zeta) \,{\rm d}\zeta
\\ \label{finalu2}
&=&  \frac{Z\alpha^2}{\gamma}
 \left( \frac{3}{16} \frac{\kappa(2\kappa -1)}{4\gamma^2-1} +
Z\alpha\,\frac{2}{3\pi}
\int_1^\infty E_1(mr_N \zeta) Y(\zeta) \,{\rm d}\zeta \right)~,
\end{eqnarray}
where  $Y(\zeta)$ is defined in (\ref{Y}) and
we used (\ref{a}),(\ref{k}) to present $a,k$ explicitly.
As is evident from (\ref{log2}) it is essential to take the 
finite size $r_N$ of the nucleus into account.
We follow in (\ref{finalu2}) the simplest way cutting
the divergent integral in (\ref{finalu2}) at $r=r_N$. 
%(later we will return to this point  
%and devise more accurate procedure, see equations (\ref{fr}),(\ref{frdfg}) 
%below).
The symbol
$E_1(x)$ in (\ref{finalu2}) 
represents the known integral-exponent function
\begin{equation}\label{E1}
E_1(x) = \int_1^\infty\exp(-xt)\frac{{\rm d}t}{t}~.
\end{equation}
%see \cite{corn}.
Formula (\ref{finalu2}) solves the problem formulated  above, giving
a  simple transparent presentation
for variation of the atomic electron wave function due to vacuum
polarization. Similarly we can find contribution of the QED vacuum
polarization to the parity-violating weak interaction.
Substituting (\ref{uehling}) into (\ref{weak}) 
and making transformations similar to the ones used in (\ref{finalu2})
we find

 \begin{equation}\label{w}
\!\!\!\!\!\!\!\!\!\!
\frac{\delta
 \langle p_{1/2}|W|s_{1/2}\rangle} {\langle p_{1/2}|W|s_{1/2}\rangle} =
  \frac{Z \alpha^2}{\gamma} \left(
\,\frac{3}{4}\,\,\frac{1}{4\gamma^2-1} +
Z\alpha \,\frac{4}{3\pi}
\int_1^\infty E_1(mr_N \zeta) Y(\zeta) \,{\rm d}\zeta \right)~.
 \end{equation}
Equation (\ref{w}) presents 
the weak interaction matrix element for an arbitrary atom in a 
transparent analytical form without fitting parameters.
Numerical results are easily obtained by a straightforward one-dimensional
integration in (\ref{w}). One only needs to specify the
nuclear size that can be taken as $r_N = 1.2 \cdot 10^{-13} A^{1/3}$ cm
where $A$ is the atomic number, see \cite{LLIV}.
\footnote{Alternatively the right-hand side of (\ref{w})
can be calculated using an expansion in powers of $mr_N \ll 1$
that reads $\frac{\alpha}{\gamma}\{ \,\frac{3}{4(4\gamma^2-1)}\,Z\alpha
+\frac{2}{3\pi}\,(Z\alpha)^2\,[\, 
(\,\ln\frac{2}{mr_N}-C-\frac{5}{6}\,)^2
+0.759\,] \, \} + O(mr_N)$, where $C\simeq 0.577$. This
expansion brings (\ref{w}) to a form that is close, but not identical to
the one derived in \cite{milstein-sushkov-01}. We will not pursue
an origin for this discrepancy since calculations in the
cited paper were fulfilled up to a constant that 
% eventually (at the end of the day)  
was eventually used as a fitting parameter.}
For the most interesting case of the $^{133}{\mathrm Cs}$ formula (\ref{w})
gives correction produced by the Uehling potential $ 0.47 \% $.

Compare this result  with other results obtained recently. 
Johnson, Bednyakov and Soff in Ref. \cite{johnson01}
calculated correction due to the Uehling potential
for the parity-nonconservation in the 6s-7s amplitude in $^{133}$Cs. 
It proves to be large $0.4\% $,
which agrees with qualitative expectations 
expressed by Sushkov in \cite{sushkov_01} previously. 
The result of \cite{johnson01} includes, along with
variation of the weak matrix element, variations of the dipole matrix
element and the corresponding energy denominator that, combined
together, describe a $s-s$ mixing measured experimentally.
Ref. \cite{dzuba01}  of Dzuba, Flambaum and Ginges 
confirmes this result and supplies more details 
providing separate variations for all
three quantities mentioned above. It was found that variations of the
dipole matrix element and the energy denominator, being not small,
compensate each other almost completely.
Thus the variation of the weak matrix
element proves to be $0.4 \% $.
Numerical calculations in \cite{milstein-sushkov-01} were restricted 
by the logarithmic accuracy that was improved 
by using a constant  as a fitting parameter
to obtain $0.4\% $ in line with \cite{johnson01}.

We are interested in heavy atoms where the parameter $Z \alpha$ is not
small, therefore the lowest order polarization potential (Uehling 
potential) may be not sufficient. The higher order polarization
potential (Wichmann-Kroll potential) was obtained in   
\cite{wichmann-kroll}.
To calculate the correction to the weak matrix element with
the logarithmic accuracy it is enough to know this potential
at small distances \cite{milstein-strakhovenko-83}:
\begin{equation}\label{wk}
V_{\rm WK}(r) = 0.092 \,\frac{2 \alpha}{3\pi}\,(Z \alpha)^2 \,
 \left( \frac{Ze^2}{r} \right) \,
  ~~~~~mr\ll 1~.
 \end{equation}
The calculation with the logarithmic accuracy gives the following
ratio of the Wichmann-Kroll correction to the Uehling correction
for the weak matrix element (see (\ref{weak})):
\begin{equation}\label{ratio}
\frac{\delta W_{\rm WK}}{\delta W_U} = -0.184 \, \frac{ (Z \alpha)^2}
{\ln \Big(1/(mr_N)\Big)}
 \end{equation}
For $^{133}$Cs  this ratio is about -0.007. This confirms 
the statement of \cite{milstein-sushkov-01} 
that the higher order corrections
to the polarization potential are not important (this may probably
be explained
by high momenta of the electron-positron pair in the polarization loop).

\section{Large separations, non-relativistic case}
\label{nonrel}
Let us apply formula (\ref{psi0})
for the case when the perturbative potential $V(r)$ is located
in a region of distances $r$ that satisfy the following conditions
\begin{equation}\label{1Z1}
a_0/Z \le r \le a_0~.
\end{equation}
An example of an application here may be the calculation of QED corrections
to the weak matrix element which originate from the atomic electron
potential (see below).

 Two simplifications are possible here. Firstly,
the motion can be described by non-relativistic
equations, and, secondly,
the semiclassical approach is valid here \cite{LLIII}.
We can therefore assume that the Dirac
spinors $\psi_{jl}({\bf r},E),\psi_{njl}({\bf r})$
can be expressed in terms of the single-component
nonrelativistic wave functions. The angular components
of the nonrelativistic wave functions
will be called $\psi_{l}(r,E)$ and $\psi_{nl}(r)$ respectively.
Applying conventional semiclassical methods \cite{LLIII}
in the classically allowed region, which includes
all distances specified in (\ref{1Z1}),
we can write
\begin{equation}\label{sc}
\psi_{l}(r,E) = \frac{1}{r}\,
\left( \frac{2}{\pi}\frac{\omega(E)}{v(r)} \right)^{1/2}
\sin \left( \int^r p(r'){\rm d}r'\right)~,
\end{equation}
where $p(r)$ and $ v(r)$ are a classical momentum and velocity,
$p(r) = mv(r)$, and $\omega(E)$ is a classical frequency
\begin{equation}\label{omega}
\omega(E) =\frac{2 \pi}{T(E)}~,~~~~~~T(E) = \oint \frac{{\rm d}r}{v(r)}~.
\end{equation}
Generally speaking,
the velocity $v(r)$ depends on the energy $E$, making
the period of the classical motion $T(E)$ and the frequency
$\omega(E)$  energy dependent as well.
Recall, however,
that we are interested in the behaviour of an outer
electron whose binding energy is much lower than an atomic
potential when $r$ satisfies (\ref{1Z1}). This fact makes
the velocity and
momentum in the integrand in
(\ref{sc})  almost independent on energy $E$ in the vicinity
of the $n$-th energy level.
In contrast, $\omega(E)$ exhibits rapid variation with energy because
the integral for the period $T(E)$ in (\ref{omega}) is saturated
at large distances, where velocity sharply depends on energy.
Taking this  into account we deduce from (\ref{sc})
that in the region of interest (\ref{1Z1}) the following
equality holds
\begin{equation}\label{scde}
\frac{\partial \psi_{l}(r,E) }{\partial E} \simeq
\frac{1}{2\omega(E)} \frac{d\omega(E)}{dE}
\,\psi_{l}(r,E)~.
\end{equation}
This shows that in the region (\ref{1Z1})
the derivative of the wave function
$\partial \psi_{l}(r,E)/\partial E$
can be described
by a simple scaling factor $(1/2\omega(E)) (d\omega(E)/dE)$.
This statement remains true for the Dirac spinor
$\partial \psi_{jl}({\bf r},E)/\partial E$ as well
because in the considered  region (\ref{1Z1})
the spinor is proportional to the nonrelativistic wave
function (\ref{scde}).
\footnote{In what follows we will not need
an explicit form for relations expressing the spinor
via the nonrelativistic wave function. A simple fact of their linear
dependence will be sufficient.}
Using now the fact that for shorter separations
$r< a_0/Z$ the perturbation is assumed insignificant, we conclude that
description of the perturbation by the scaling factor
remains valid all the way down to the nucleus.
This means that for
all distances inside an atomic core $0\le r \le a_0$ the derivative of the
Dirac spinor over energy $\partial \psi_{jl}({\bf r},E)/\partial E$
remains proportional to the spinor itself $\psi_{jl}({\bf r},E)$ with a
scaling coefficient identical to the one in the right-hand side of
(\ref{diffnorma}). Using this result in (\ref{psi0}) we find
\begin{equation}\label{omEom}
 \delta \psi_{njl}({\bf r}) =
\frac{\omega'(E_n)}{2 \omega(E_n)}\,\psi_{njl}({\bf r}) \delta E_n
\end{equation}
Here $\omega'(E)$ is a shortcut notation for a derivative of $\omega(E)$
over $E$, while $\delta E_n$ is an energy variation due to the potential
$V(r)$
\begin{equation}\label{deltaE} \delta E_n =
\langle njl| V | njl \rangle \equiv
\int \psi_{njl}^+({\bf r})V(r) \psi_{njl}({\bf r}) \,{\rm d} ^3r~.
\end{equation}
We conclude from (\ref{omEom}) that the relative variation
of the wave function at the origin is proportional to the
variation of the energy level,
being independent on any  specific features of the potential.
The coefficient $\omega'(E_n)/(2\omega(E_n))$ in this
formula is expressed in terms of the classical frequency for the
electron motion. It is very simple for calculations, but can
be simplified even further, if one needs only an estimation.
Remember again that
large separations from the atom $r>a_0$ give large contribution to the
classical period $T(E)$. For these distances an atomic field can be
approximated by the Coulomb potential $-e^2/r$ created by a singly charged
atomic residue.  This fact allows one to approximate the frequency by
conventional formula of Newtonian celestial mechanics for the Kepler
problem which for the potential $-e^2/r$ read
$\omega(E) =
(\pi/e^2)(2B_n^3/m)^{1/2}$, where $B_n=m-E_n$ is the electron binding
energy.  Substituting this in (\ref{omEom}) one finds
\begin{equation}\label{kep}
\delta \psi_{njl}(0) \simeq \frac{3}{4}
\frac{\delta B_n}{B_n}\,\psi_{njl}(0)~,
\end{equation}
where $\delta B_n = m-\delta E_n$ is the variation of the
binding energy.
Thus the behaviour of the wave function can be  described in terms of the
binding energy only.
\footnote{It is amusing to observe that the coefficient in the
right-hand side of (\ref{kep}) originates directly from the Kepler law that
relates cubes of periods with squares of separations.} There is, of course,
a short-cut way to derive this result.  The wave function of an outer
electron at small $r$ is known to depend on the binding energy according to
$\psi_{nl}^2(r) = const/\tilde n^3$,
where $\tilde n$ is an effective radial quantum number
defined by the binding energy $B_n = m e^4/(2\tilde n^2)$.  Taking
variation of this relation and assuming a weak influence
of the perturbation on $const$ 
one immediately reproduces (\ref{kep}).

The numerical simulation performed
\footnote{An atomic potential was approximated by some local
potential which {\em reasonably} reproduces the valence electron
wave function both inside and outside the atomic core.}
shows that for the $6s_{1/2}$ state in cesium atom
the Kepler approximation (\ref{kep}) ensures an accuracy of
$\sim 20\%$ for all perturbative
potentials $V(r) = const \cdot\exp(-pr)$ with
$1/a_0 \le p \le 50/a_0 $, while an accuracy of
slightly more sophisticated formula (\ref{omEom}) is even higher,
of the order of $2\%$.

We verified in this Section
that the approach based on (\ref{psi0}) gives sensible results
(\ref{omEom}),(\ref{kep}) for the region (\ref{1Z1}).
Now we can use this approach to estimate the influence
of QED radiative corrections to electron-electron interaction
on the weak matrix element. In the non-relativistic limit
the Uehling potential can  be replaced by a zero-range
potential (proportional to the $\delta$-function). 
A larger correction comes from the self-energy
operator $\Sigma ({\bf r},{\bf r}',E)$ which also reduces to 
 $\delta$-function in the non-relativistic limit, i.e.
$\Sigma ({\bf r},{\bf r}',E) 
\sim \delta({\bf r}-{\bf r}')\nabla^2 U({\bf r})$.   
A semiclassical formula for $\delta U({\bf r})$ 
was obtained by Flambaum and Zelevinsky \cite{flambaum-zelevinsky-99}
\begin{equation}
\label{deltaU}
 \delta U({\bf r})= \frac{ Z^2 \alpha}{3 \pi m^2} 
\ln\frac{m}{|U_{p}({\bf r})-E|} \nabla^2 U({\bf r}).
\end{equation}
Here $U({\bf r})$ is the atomic potential, 
while $U_p({\bf r})$ is the atomic potential with a correction
that takes into account  the centrifugal potential which influence
the $p$-wave electron in the intermediate state 
(see details in \cite{flambaum-zelevinsky-99}). 
As usual, this semiclassical expression is not valid near the
turning points where $U_{p}(r)=E$. However, a very weak logarithmic singularity
does not produce any practical limitations on the applicability
of (\ref{deltaU}). For the electrostatic potential
$ \nabla^{2}U({\bf r}) = - 4 \pi  \rho (r)$ where  $\rho (r)$
is the electric charge density. The main contribution to the Lamb shift
of the energy level produced  by $\delta U({\bf r})$ is given
by the nuclear charge (this contribution was calculated 
in \cite{dzuba01}). The contribution of the electron charge density
can be calculated using the Thomas-Fermi approximation. 
The main contribution here comes from the interval  $a_0/Z < r <a_0/Z^{1/3}$.
A simple estimate shows that the squared  semiclassical electron wave
 function at  $r \sim a_0/Z^{1/3}$ is $Z$ times smaller than near
 the origin. Therefore, the electron-electron contribution
to the Lamb shift of s-wave electron is $Z$ times smaller than the
 electron-nucleus contribution. According to \cite{dzuba01} the Lamb shift
of the s-levels in $^{133}$Cs  is $\sim 0.1\%$ . Then using equation (\ref{kep}) 
we obtain that the correction to  the weak matrix element
produced by the electron density contribution to the  $\delta U({\bf r})$
is $\sim 0.001 \%$.

 \section{Summary and conclusions}
 Equation (\ref{psi0}) provides a convenient
 framework to calculate
 a variation of the wave function of an atomic electron inside
the atomic nucleus that arises due to a perturbative potential
 in an atom. Applied to
 the region of {\em large} distances ($a_0/Z\le r \le a_0$) it
results in equation (\ref{omEom}) and its simplified version
 (\ref{kep}).  There is a reason that makes them interesting
 for applications for the atomic parity nonconservation. 
 The most
 difficult and cumbersome part of theoretical investigation
in the latter problem present many-electron
 correlations \cite{dzuba89,blundell90,blundell92,kozlov01,dzuba01}.
 The correlations take place exactly in the
 region of large distances discussed in Section \ref{nonrel}.
 We can deduce from
 (\ref{omEom}),(\ref{kep}) that an accuracy of calculations of the atomic
 spectrum provides a  direct test for an accuracy of the
 weak matrix element calculation. 
 There are, of course,  other problems  which require knowledge of
an electron wave function in the vicinity of a nucleus such as
hyperfine interaction, field isotopic shift,  and
time invariance violation.
 
 In the region of {\em small} separations $r \ll a_0/Z$ our main result
 (\ref{final}) gives simple transparent
 presentation for the influence of perturbation on the wave function.
 We deduce from it that for potentials singular at the origin a
 dimensionless parameter $ ma \int V(r){\rm d} r$  (\ref{a})
 measures the strength of the potential.

 Formula (\ref{final}) was applied to find a variation of the
 weak electron-nucleus matrix element due to the vacuum
 polarization (Uehling potential).
 The result is $0.47 \% $ for $^{133}$Cs atom, which
 agrees with the results reported recently in 
\cite{johnson01,milstein-sushkov-01,dzuba01}.
The contributions of the Wichmann-Kroll potential and QED corrections
to the electron-electron interaction were found to be very small.

\ack
We are grateful to M.Kozlov, A.Milstein and O.Sushkov for  
useful discussions.
This work was supported by the Australian Research Council.
%************************************************************

 %\begin{literature}
 %\end{literature}

  \end{document}